\documentstyle[12pt]{article}

\catcode`\@=11
\long\def\@makefntext#1{ 
\protect\noindent \hbox to 3.2pt {\hskip-.9pt
$^{{\ninerm\@thefnmark}}$\hfil}#1\hfill} 

\def\thefootnote{\fnsymbol{footnote}}
 \def\@makefnmark{\hbox to 0pt{$^{\@thefnmark}$\hss}}  

\def\ps@myheadings{\let\@mkboth\@gobbletwo
\def\@oddhead{\hbox{} 
\rightmark\hfil\ninerm\thepage}
\def\@oddfoot{}\def\@evenhead{\ninerm\thepage\hfil 
\leftmark\hbox{}}\def\@evenfoot{}
\def\sectionmark##1{}\def\subsectionmark##1{}}

\begin{document}

\newcommand{\symbolfootnote}{\renewcommand{\thefootnote}
	{\fnsymbol{footnote}}}
\renewcommand{\thefootnote}{\fnsymbol{footnote}}
\newcommand{\alphfootnote}
	{\setcounter{footnote}{0}
	 \renewcommand{\thefootnote}{\sevenrm\alph{footnote}}}

\newcounter{sectionc}\newcounter{subsectionc}\newcounter{subsubsectionc}
\renewcommand{\section}[1] {\vspace{0.6cm}\addtocounter{sectionc}{1}
\setcounter{subsectionc}{0}\setcounter{subsubsectionc}{0}\noindent
	{\bf\thesectionc. #1}\par\vspace{0.4cm}}
\renewcommand{\subsection}[1] {\vspace{0.6cm}\addtocounter{subsectionc}{1}
	\setcounter{subsubsectionc}{0}\noindent
\renewcommand{\subsubsection}[1] {\vspace{0.6cm}
\addtocounter{subsubsectionc}{1}
	\noindent {\rm\thesectionc.\thesubsectionc.\thesubsubsectionc.
	#1}\par\vspace{0.4cm}}
\newcommand{\nonumsection}[1] {\vspace{0.6cm}\noindent{\bf #1}
	\par\vspace{0.4cm}}

\newcounter{appendixc}
\newcounter{subappendixc}[appendixc]
\newcounter{subsubappendixc}[subappendixc]
\renewcommand{\thesubappendixc}{\Alph{appendixc}.\arabic{subappendixc}}
\renewcommand{\thesubsubappendixc}
	{\Alph{appendixc}.\arabic{subappendixc}.\arabic{subsubappendixc}}

\renewcommand{\appendix}[1] {\vspace{0.6cm}
        \refstepcounter{appendixc}
        \setcounter{figure}{0}
        \setcounter{table}{0}
        \setcounter{equation}{0}
        \renewcommand{\thefigure}{\Alph{appendixc}.\arabic{figure}}
        \renewcommand{\thetable}{\Alph{appendixc}.\arabic{table}}
        \renewcommand{\theappendixc}{\Alph{appendixc}}
        \renewcommand{\theequation}{\Alph{appendixc}.\arabic{equation}}
       \noindent{\bf Appendix \theappendixc. #1}\par\vspace{0.4cm}}
        \noindent{\bf Appendix \theappendixc #1}\par\vspace{0.4cm}}
\newcommand{\subappendix}[1] {\vspace{0.6cm}
        \refstepcounter{subappendixc}
        \noindent{\bf Appendix \thesubappendixc.
#1}\par\vspace{0.4cm}} \newcommand{\subsubappendix}[1] {\vspace{0.6cm}
        \refstepcounter{subsubappendixc}
        \noindent{\it Appendix \thesubsubappendixc. #1}
	\par\vspace{0.4cm}}

\def\abstracts#1{{
	\centering{\begin{minipage}{30pc}\tenrm\baselineskip=12pt\noindent
	\centerline{\tenrm ABSTRACT}\vspace{0.3cm}
	\parindent=0pt #1
	\end{minipage} }\par}}

\newcommand{\bibit}{\it}
\newcommand{\bibbf}{\bf}
\renewenvironment{thebibliography}[1]
	{\begin{list}{\arabic{enumi}.}
	{\usecounter{enumi}\setlength{\parsep}{0pt}
\setlength{\leftmargin 1.25cm}{\rightmargin 0pt}
	 \setlength{\itemsep}{0pt} \settowidth
	{\labelwidth}{#1.}\sloppy}}{\end{list}}
\def\np{{\em Nucl. Phys.}}
\def\pl{{\em Phys. Lett.}}
\def\pr{{\em Phys. Rev.}}
\def\prl{{\em Phys. Rev. Lett.}}

\topsep=0in\parsep=0in\itemsep=0in
\parindent=1.5pc

\newcounter{itemlistc}
\newcounter{romanlistc}
\newcounter{alphlistc}
\newcounter{arabiclistc}
\newenvironment{itemlist}
    	{\setcounter{itemlistc}{0}
	 \begin{list}{$\bullet$}
	{\usecounter{itemlistc}
	 \setlength{\parsep}{0pt}
	 \setlength{\itemsep}{0pt}}}{\end{list}}

\newenvironment{romanlist}
	{\setcounter{romanlistc}{0}
	 \begin{list}{$($\roman{romanlistc}$)$}
	{\usecounter{romanlistc}
	 \setlength{\parsep}{0pt}
	 \setlength{\itemsep}{0pt}}}{\end{list}}

\newenvironment{alphlist}
	{\setcounter{alphlistc}{0}
	 \begin{list}{$($\alph{alphlistc}$)$}
	{\usecounter{alphlistc}
	 \setlength{\parsep}{0pt}
	 \setlength{\itemsep}{0pt}}}{\end{list}}

\newenvironment{arabiclist}
	{\setcounter{arabiclistc}{0}
	 \begin{list}{\arabic{arabiclistc}}
	{\usecounter{arabiclistc}
	 \setlength{\parsep}{0pt}
	 \setlength{\itemsep}{0pt}}}{\end{list}}

\newcommand{\fcaption}[1]{
        \refstepcounter{figure}
        \setbox\@tempboxa = \hbox{\tenrm Fig.~\thefigure. #1}
        \ifdim \wd\@tempboxa > 6in
           {\begin{center}
        \parbox{6in}{\tenrm\baselineskip=12pt Fig.~\thefigure. #1 }
            \end{center}}
        \else
             {\begin{center}
             {\tenrm Fig.~\thefigure. #1}
              \end{center}}
        \fi}

\newcommand{\tcaption}[1]{
        \refstepcounter{table}
        \setbox\@tempboxa = \hbox{\tenrm Table~\thetable. #1}
        \ifdim \wd\@tempboxa > 6in
           {\begin{center}
        \parbox{6in}{\tenrm\baselineskip=12pt Table~\thetable. #1 }
            \end{center}}
        \else
             {\begin{center}
             {\tenrm Table~\thetable. #1}
              \end{center}}
        \fi}

\def\@citex[#1]#2{\if@filesw\immediate\write\@auxout
	{\string\citation{#2}}\fi
\def\@citea{}\@cite{\@for\@citeb:=#2\do
	{\@citea\def\@citea{,}\@ifundefined
	{b@\@citeb}{{\bf ?}\@warning
	{Citation `\@citeb' on page \thepage \space undefined}}
	{\csname b@\@citeb\endcsname}}}{#1}}

\newif\if@cghi
\def\cite{\@cghitrue\@ifnextchar [{\@tempswatrue
	\@citex}{\@tempswafalse\@citex[]}}
\def\citelow{\@cghifalse\@ifnextchar [{\@tempswatrue
	\@citex}{\@tempswafalse\@citex[]}}
\def\@cite#1#2{{$\null^{#1}$\if@tempswa\typeout
	{IJCGA warning: optional citation argument
	ignored: `#2'} \fi}}
\newcommand{\citeup}{\cite}

\def\fnm#1{$^{\mbox{\scriptsize #1}}$}
\def\fnt#1#2{\footnotetext{\kern-.3em
	{$^{\mbox{\sevenrm #1}}$}{#2}}}

\font\twelvebf=cmbx10 scaled\magstep 1
\font\twelverm=cmr10 scaled\magstep 1
\font\twelveit=cmti10 scaled\magstep 1
\font\elevenbfit=cmbxti10 scaled\magstephalf
\font\elevenbf=cmbx10 scaled\magstephalf
\font\elevenrm=cmr10 scaled\magstephalf
\font\elevenit=cmti10 scaled\magstephalf
\font\bfit=cmbxti10
\font\tenbf=cmbx10
\font\tenrm=cmr10
\font\tenit=cmti10
\font\ninebf=cmbx9
\font\ninerm=cmr9
\font\nineit=cmti9
\font\eightbf=cmbx8
\font\eightrm=cmr8
\font\eightit=cmti8


       {\normalsize \hfill
       \begin{tabbing}
       \`\begin{tabular}{l}
         IC/95/349  \\
         hep--th/9510164 \\
         October 23, 1995 \\
         \end{tabular}
       \end{tabbing} }\vspace{8mm}

\centerline{\twelvebf SUPERSYMMETRY BREAKING IN ORBIFOLD}
\centerline{\twelvebf  COMPACTIFICATIONS}
\vspace{0.8cm}
\centerline{\tenrm KARIM BENAKLI\footnote{E-mail:
benakli@ictp.trieste.it}}
\centerline{\tenit  International Centre of Theoretical Physics,
Trieste, Italy}
\baselineskip=12pt
\vspace{0.9cm}

\renewcommand{\arraystretch}{2.0}
\renewcommand{\thefootnote}{\alph{footnote}}
\newcommand{\be}[3]{\begin{equation}  \label{#1#2#3}}
\newcommand{\ee}{\end{equation}}
\newcommand{\ba}{\begin{array}}
\newcommand{\ea}{\end{array}}
\newcommand{\vsf}{\vspace{5mm}}
\newcommand{\NP}[3]{{\em Nucl. Phys.}{ \bf B#1#2#3}}
\newcommand{\PRD}[2]{{\em Phys. Rev.}{ \bf D#1#2}}
\newcommand{\MPLA}[1]{{\em Mod. Phys. Lett.}{ \bf A#1}}
\newcommand{\PL}[3]{{\em Phys. Lett.}{ \bf B#1#2#3}}
\newcommand{\marpar}{\marginpar[!!!]{!!!}}
\newcommand{\lab}[2]{\label{#1#2}   (#1#2) \hfill }

\abstracts{
 Known mechanisms for breaking of supersymmetry at
the level of string theory imply that at least one of the internal
dimensions has a very large size. Experimental detection of the
associated light Kaluza-Klein (KK) excitations would be a
strong hint for the existence of string like elementary objects, as no
consistent field theory describing them is known. We restrict the
discussion to the Scherk-Schwarz mechanism in
orbifold compactifications. For this case we investigate the quantum
number of the lightest predicted KK states.}
\vspace{10mm}

\vfill
\baselineskip=14pt

\begin{center}
Talk presented at the \\
{\em Workshop on strings, gravity and related topics} \\
Trieste, Italy, June 29 -- 30, 1995
\end{center}

\vfill

\newpage

\twelverm   
\baselineskip=14pt

Supersymmetry appears quite naturally in superstring theory. However
understanding its breaking remains an open problem. If this
breaking is due to non-perturbative effects then it can not be
studied directly at the level of string models within the
actual perturbative formulation. Another possibility is that
supersymmetry is broken at tree level. This is the case of
supersymmetry breaking by a magnetic field\cite{Bac} or
through the string version of the Scherk-Schwarz mechanism\cite{SSb}.
In both case one finds that the gravitino or gauginos get masses
inversely proportional to the size of some internal dimension. This is
in agreement with the result that a small supersymmetry breaking scale
implies a large internal dimension\cite{Pre}. Here we review the case
of the Scherk-Schwarz mechanism where the fermions and bosons have
mass splitting due to different compactification boundary
conditions\cite{SS}.

The simplest framework to study supersymmetry breaking through
the Scherk-Schwarz mechanism are the orbifold
compactifications\cite{orb}. These are four-dimensional string models
with $N=1$ space-time supersymmetry obtained from toroidal
compactification by dividing out some discrete subgroup of the
automorphisms of the Hilbert space. Here the elements of the discrete
subgroup are combinations of translation shifts and  rotations. The
resulting physical Hilbert space consists in twisted and untwisted
sectors. The twisted sectors contain states that don't have internal
momenta so, at the tree level, they don't feel the supersymmetry
breaking mechanism. The untwisted sector is obtained from the Hilbert
space of a string  propagating on a torus by projecting on invariant
states under the action of the orbifold. The mass spectrum in this
sector is determined by the associated internal momenta
through\cite{Nar}$^-$\cite{KB}:

\begin{equation}
{1\over 4}m_L^2={1\over 4}m_R^2=N_R + {1\over 2} {\bf p}_R^2 ,
\label{eq:thirty}
\end{equation}
where $N_R$ is the oscillator
number and  ${\bf p}_R$ is the internal momentum given by:

\begin{equation}
p_R = ( m_i - a^I_i (p^I -{1\over2} a^I_j n^j)+\xi^*_{ij}
Q^j - {{\xi_{ki} \xi^*_{kj}}\over 2} n^j) {{\bf e}^{*i} \over {2R_i} }
- n^i R_i  {\bf e}_i.
\label{eq:pr}
\end{equation}

In the above formula $p^I$ is the gauge internal momentum, $m_i$
are the momenta number, $n^i$ are winding number. $a^I_i$ are the
Wilson lines and  $Q^j$ is the charge that takes integer and
fractional values for the bosons and fermions respectively, breaking
supersymmetry. The requirement that the orbifold projection and gauge
symmetry breaking commute imposes a condition on the allowed Wilson
lines\cite{{orb},{Wil}} and reduces the maximum number of
independent discrete Wilson lines. The parameters $\xi^*_{ij}$ take
discrete values and they parameterize the Lorentz boost which takes
the theory from the unbroken supersymmetric phase to the broken one.
Moreover the gauge symmetry breaking is also achieved through a
Lorentz boost\cite{{Nar},{Hig}}.  The
combination of the both Wilson lines and Scherk-Schwarz charge is then
equivalent to a boost on the vector $(Q^A, p^I, p^a_L; p^{a'}_R)$
given by\cite{KB}:

\begin{equation}{\left(\matrix{{\delta_{AB}}&0&{-{1\over2}\xi_{Ab} }
&{{1\over2}\xi_{Ab'} }\cr
0&{\delta^{I}_{J}}&{{1\over2}A^I_{b}}
&{-{1\over2}A^I_{b'}}  \cr
{{1\over2}\xi^*_{aB} }&{-{1\over2}A^J_a}&{\delta_{ab}-{1\over
8}\xi_{Ca} \xi^*_{Cb}
 -{1\over8} A^{K}_aA^{K}_{b}}
&{1\over8}{\xi_{Ca} \xi^*_{Cb'}
 +{1\over8}
A^{K}_aA^{K}_{b'}}\cr
{{1\over2}\xi^*_{a'B} }&{-{1\over2}A^J_{a'}}&{- {1\over8}\xi_{Ca'}
\xi^*_{Cb} -{1\over8} A^{K}_{a'} A^{K}_{b}}&
{\delta_{a'b'}+{1\over8}\xi_{Ca'} \xi^*_{Cb'}+{1\over8} A^K_{a'}
A^{K}_{b'}}\cr}\right)} \label{eq:boost}
\end{equation}
which leads to the spectrum (\ref{eq:pr}).

It is important to notice that the breaking of gauge symmetry and
supersymmetry commute as they are two similar (but different)
Lorentz boosts. Then there is no new condition imposed on the charge
$Q^A$. That allows us to study the properties of the models in their
supersymmetric phase.

The requirement that orbifold projection, gauge symmetry and
supersymmetry breaking commute, restricts the allowed Wilson lines and
the charges $Q^A$. The charge $Q^A$ can be written as:
$Q^A= \oint J^A$ where $J^A$ is a $U(1)$ current which shouldn't
commute with the $2d$ supercurrent so that it gives a charge for the
gravitino but not to the graviton and gauge bosons which are usually
in the untwisted sector.

In orbifold compactifications, the condition that $Q^A$ is associated
with some particular direction $A$ means that it should have the same
transformation under the orbifold group than the corresponding
coordinate $\partial X^A$. This requirement is very strong as it leaves
only few possible $U(1)$ currents.

Different currents were found for the cases of $Z_N$ and $Z_N\times
Z_M$ orbifolds\cite{KB}. For the orbifolds ${\bf Z}_4$ and
${\bf Z}_2 \times {\bf Z}_2$ the $U(1)$ charges were already
known\cite{Ant}. We have focused on orbifolds where only one dimension
(${\bf Z}_2$ case) or two dimensions are large. For example, this
excludes ${\bf Z}_7$ orbifold which needs the six internal dimensions
to be of the same size. For the orbifolds ${\bf Z}_3$, ${\bf Z}_6$,
${\bf Z}_8$ and ${\bf Z}_{12}$ as well as ${\bf Z}_2 \times {\bf Z}_6
$, ${\bf Z}_3 \times {\bf Z}_3$,  ${\bf Z}_3 \times {\bf Z}_6$, ${\bf
Z}_4 \times {\bf Z}_4 $ and ${\bf Z}_6 \times {\bf Z}_6$ we have not
found charges allowing to implement the Scherk-Schwarz mechanism.

Such theories with perturbative breaking of supersymmetry have become
recently of some phenomenological interest after it has been shown
that they could allow a weakly coupled string theory, at least at
one-loop for a class of models based on orbifold
compactifications\cite{Ant}. We have found only two orbifolds ${\bf
Z}_4$ and ${\bf Z}'_6 \equiv {\bf Z}_2 \times {\bf Z}_3$ which have
charges associated with $N=4$ sectors where the possible large
threshold corrections vanish. The other orbifolds lead to light
KK-states in $N=2$ multiplets. In this case, the one loop threshold
depending on the value of the large radius is not automatically
vanishing. One would have then to chose the particle content as
KK-excitations to get vanishing $\beta$-functions\cite{Ant}.

In these theories the manifestation of the large extra-dimension(s)
would be the existence of KK-excitations that would appear as some new
particles with regularly spaced masses and behaving as excitations of
the MSSM particles. In the limit where some supersymmetry
(thus electroweak) breaking effects are neglected, some properties
as the quantum numbers and interactions of these states in orbifold
compactifications have been investigated\cite{AB}. The
viability of these theories requires that\cite{Aqm}:

i) As the fermions from the untwisted sector acquire a
common mass-shift, the quarks and leptons must be identified with
twisted states. This rules out all the
string $SU(3)\times SU(2) \times U(1)$ models build in the past.

ii) If the Higgs doublets appear in the untwisted sector, the
generation of masses for untwisted fermions allows for generating
the $\mu$- term.

It has been pointed out\cite{AB} that if the untwisted sector is
`minimal' the only observable effects of the KK-excitations are through
some non-renormalizable effective operators. The latter have then been
computed and limits on the size of new dimensions have been derived
from actual experimental data\cite{AB}. The obtained bounds allow the
hope of experimental detection in the near future\cite{ABQ}. They
could also have some cosmological implications\cite{AS}.

An obvious question to address is the compatibility of these
requirements with orbifold compactifications. In other words, what are
the new light states we expect in realistic models?. Notice first that
in the gauge symmetry breaking process, the states acquire masses
inversely proportional to the radius of the torus corresponding to the
Wilson line. The massless states (in the supersymmetric phase) can
easily be seen to correspond to Wilson lines singlets: $aP \in {\bf
Z}$. While we see that a Wilson line associated to the torus with the
large radius used to break supersymmetry will lead to a mass of the
order of hundreds GeV or TeV to the projected states. In particular,
if some states have $0 < |aP| < 1$ as it is often the case, then the
corresponding states will have masses smaller than the KK excitations
of the states, with different gauge quantum numbers, present at the
massless level. This also implies that the minimal light KK states are
obtained when all the Wilson lines have to be associated only to the
other small tori. Such a minimally requirement would also
automatically avoid the presence of some massive new vector bosons
that could mediate new dangerous interactions.

The formula (\ref{eq:pr})  shows that all the states
carrying the same gauge internal momenta have the same masses. In
particular, this implies that all the $N=2$ and $N=4$ multiplets get
projected by the gauge symmetry breaking and only $N=4$ (or $N=2$)
excitations of massless untwisted states are present among the light
KK states.

We have also to deal with the effect of reducing the
rank of the gauge group on the Kaluza-Klein excitations. The Higgs
mechanism through discrete Wilson lines described above doesn't reduce
the rank of the gauge group. To reduce the rank one can embed the
Wilson lines in the gauge group as automorphism of the $\Gamma_{16}$
lattice\cite{rank}. This corresponds to the case where the orbifold
action on the gauge lattice is described by a rotation ${\Theta}\neq
0$. In this case some Cartan generators of the gauge group are not
associated with a root of $\Gamma_{16}$, but with an invariant
combination of winding states. In the case where some components of
the Wilson line are rotated by ${\Theta}$, the
projection on Wilson line singlets projects out, in general, the
Cartan generators which have the form of invariant combination of
winding states. As this projection is at the level of the gauge
lattice state, which is the same for all the KK excitations of
the gauge boson, all the KK tower is projected out. Both the rank of
the gauge symmetry group and the rank of the symmetry group of the
KK excitations are reduced simultaneously. Notice that if the Wilson
line is associated with the dimension with large size then the
projected states are very light.

We have investigated\cite{KB2} the minimal light untwisted states
obtained from ${\bf Z}_4$ and ${\bf Z}'_6 \equiv {\bf Z}_2 \times
{\bf Z}_3$ orbifolds. For ${\bf Z}_4$ we found that from $E_8 \times
E_8$, the minimal untwisted spectrum is the adjoint representation of
$SU(2)\times SU(4) \times U(1)\times \cdots$ or $SU(4)\times SU(4)
\times U(1)\times \cdots$ if we want Higgs like doublets from the
untwisted sector. From $Spin(32)/Z_2$ we can obtain $SU(3)\times SU(3)
\times U(1)\times \cdots$ by using for example the following Wilson
lines: \begin{eqnarray} a_1&={1\over 4}( \ \ 1, \ 1,-1,-1,-1,-1, \ \
1, \ \ 1,  \ \ 1, \ \ 1,\ \ 1,\ \ 1,\ \ 1,\ \ 1,\ \ 1,\ \ 1);\\[3mm]
 a_2&={1\over 4}( \ \ 0, \ \ 0,\ \ 2,
\ \  2, \ \ 2,\ \ 2, \ \ 0, \ \ 0, \ \ 0, \ \ 0,\ \ 0, \ \ 0,
\ \ 0,\ \ 0, \ \ 0, \ \ 0);\\[3mm] a_3&={1\over 4}( \ \ 2, \ \ 2, \ \ 0, \
\ 0,
\ \ 0, \ \ 2, \ \ 2, \ \ 0, \ \ 0, \ \ 0, \ \ 0, \ \ 0, \ \ 0, \ \ 0,
\ \ 0,  \ \ 0);
\end{eqnarray}
but it is hard to find corresponding orbifold shifts leading to chiral
quark doublets in the spectrum. For ${\bf Z'}_6$ we
found that from $E_8 \times E_8$ we can get $SU(3)\times SU(3) \times
U(1)\times \cdots$ but as for the first case it is (at least) difficult
to get three generation models.

Many questions regarding the proposed mechanism for breaking SUSY
still remain open. However one of the nice features of this scenario is
that it makes precise predictions that could be tested at future
colliders.

\vskip 1.0 cm
\noindent{\bf Acknowledgements}
\vskip 0.5 cm
I wish to thank I. Antoniadis, E. Gava, K.S. Narain, M. Quir\`os, J.
Rizos and A. Sen for discussing different parts of the material
presented here.

\end{document}